\documentclass[iop,tighten,apj]{emulateapj}

\slugcomment{Accepted for publication in ApJ}

\begin{document}

\title{Effect of dark matter halo substructures on galaxy rotation curves}

\author{Nirupam Roy}
\affil{NRAO, P. O. Box O, 1003 Lopezville Road, Socorro, NM 87801, USA}
\email{nroy@aoc.nrao.edu}

\begin{abstract}
The effect of halo substructures on galaxy rotation curves is investigated 
in this paper using a simple model of dark matter clustering. A dark matter 
halo density profile is developed based only on the scale free nature of 
clustering that leads to a statistically self-similar distribution of the 
substructures at galactic scale. Semi-analytical method is used to derive 
rotation curves for such a clumpy dark matter density profile. It is found that 
the halo substructures significantly affect the galaxy velocity field. Based 
on the fractal geometry of the halo, this self-consistent model predicts an 
NFW-like rotation curve and a scale free power spectrum of the rotation 
velocity fluctuations. 
\end{abstract}

\keywords{dark matter --- galaxies: dwarf --- galaxies: general --- galaxies: ISM --- galaxies: structure --- ISM: kinematics and dynamics}

\section{Introduction}
\label{sec:intro}

Observational and theoretical studies of galaxy rotation curves 
\citep{rubin70}, velocity dispersions of elliptical galaxies \citep{faber76}, 
baryon fraction in clusters \citep{clowe06,mass07}, gravitational lensing, 
structure formation, CMB power spectrum etc. indicate that a significant 
fraction of the mass of the Universe has no electromagnetic interaction 
\citep{bertone05}. At galactic scale, the main evidence for the existence of 
such gravitating matter with very high mass to light ratio comes from 
observations of galaxy rotation curves. The gravitational potential derived 
from the observed rotation curve can not be explained only by the visible mass 
with any reasonable mass to light ratio 
\citep{rubin70,rubin80,sofue96,sofue01,spano08}. Though there are modified 
theories of the gravitation and other suggestions 
\citep{milgrom83,beken84,sand86,fahr90,sanders97,brown06} to explain this 
anomaly, the dark matter concept is widely accepted to be a simpler 
explanation of these observations. Though observations of galaxy rotation 
curves established the existence of the dark matter at galactic scale, there 
is no general agreement on the nature and various properties (like mass 
distribution) of this major constituent of the Universe \citep{nfw96,blok05}. 
Potentially, galaxy rotation curves can also shed light on some aspects of the 
dark matter properties and thus help us in deriving a better understanding of 
the nature of the dark matter. One such key aspect is the density profile or 
the mass distribution of the dark matter at galactic scale.

On the theoretical front, there are various mass distribution models of the 
dark matter halo. The isothermal profile, the Navarro-Frenk-White (NFW) 
profile and some variant of these two profiles are used extensively in both 
theoretical and observational studies. An almost flat observed rotation curve 
outside the core of a galaxy led to the dark matter halo model with $1/r^2$ 
isothermal density profile \citep{bege91,burk95}. The density at the centre 
is not finite in this model. However, it is possible to impose appropriate 
boundary conditions to derive a non-singular solution with density 
$\rho(r)_{NIS} = \rho_0r_c^2/(r_c^2+r^2)$, where $\rho_0$ is the central 
density and $r_c$ is the ``core radius''. At large radius, this profile will 
be similar to $1/r^2$ singular isothermal profile. A cut-off radius 
$r_{max}$ is also required to be imposed to keep the total mass of the halo 
finite. However, the non-singular isothermal profile was not very successful 
in explaining the observed rotation curves. A significant progress in the 
field was the introduction of an alternative density profile, the NFW 
density profile, $\rho(r)_{NFW} = \rho_0r_c^3/[r(r_c+r)^2]$, where $\rho_c$ 
and $r_c$ are the characteristic density and characteristic radius 
respectively \citep{nfw96,nfw97,jing00}. 

The NFW profile has been quite successful in explaining the observed rotation 
curves for many galaxies. Over the time, a class of variant of the isothermal 
and the NFW profiles are also introduced to fine tune the agreement of theory 
and observations \citep{burk95,zhao96,fuku01}. All these density profiles are 
of more or less similar merit, as far as explaining the rotation curve is 
concerned. However, detailed analysis shows that the NFW profile has a number 
of problems. For example, observations of galaxies and clusters suggest a 
central flat density core \citep{blok07,kdn08}, whereas, in NFW profile, dark 
matter density has a central cusp with a logarithmic slope $\approx -1$. 
Numerical simulations show that the presence of baryon can change the mass 
distribution significantly \citep[][and references therein]{ez04,sm10}, and 
for isothermal cusp, minor merger and dynamical friction may lead to a 
shallower central density slope \citep{rd08}. The NFW profile also seems not 
to fit the observed rotation curves of the dark matter dominated low surface 
brightness galaxies and low mass dwarf galaxies \citep{blok02,blok05}. A more 
serious issue with the NFW profile is that the profile is derived by fitting 
analytical function to the density distribution derived from numerical 
simulation of structure formation with dark matter. This, in a sense, lacks a 
proper physical understanding. Though some physical insight of the NFW profile 
have recently come from further high resolution dark matter simulations 
\citep[e.g.,][]{taylor01}, the issue is far from being settled. Moreover, even 
if these cosmological simulations are very successful on large scale, at 
galactic scale there are unsolved issues, like the ``angular momentum 
problem'' \citep{sl99,sl01,burk04} or the ``missing satellite problem'' 
\citep{klypin99,moore99}, which are yet to be addressed.

Here, we have developed a self-consistent model of the dark matter halo 
substructure distribution at galactic scale to explain observed NFW-like 
rotation curves. Support for the existence of dark matter substructures has 
mainly come from numerical simulations \citep{gc08,mdk08,vs08,etws09,lad09}. 
But, there are strong observational hints like flux anomalies and time delays 
in gravitational lensing \citep{chen09,km09,vk09,xu09}, or enhanced gamma rays 
and leptonic cosmic rays \citep{ep09,pinz09}, indicating the presence of 
substructures. The present model is based on the assumption of a scale free 
nature of the dark matter clustering that leads to a statistically 
self-similar distribution of the halo substructures at galactic scale. It is 
shown that a simple fractal model of the dark matter halo substructure 
predicts an NFW-like rotation curve. Such a model also predicts a scale free 
power spectrum of the rotation velocity fluctuations. The model is described 
in Section~\ref{sec:model}, and the results are presented in 
Section~\ref{sec:result}. Possible limitations of our analysis are discussed 
in Section~\ref{sec:disc}. Finally, we summarize and present our conclusions 
in Section~\ref{sec:concl}.

\section{The Model}
\label{sec:model}

\begin{figure*}
\begin{center}
\includegraphics[width=6.25cm]{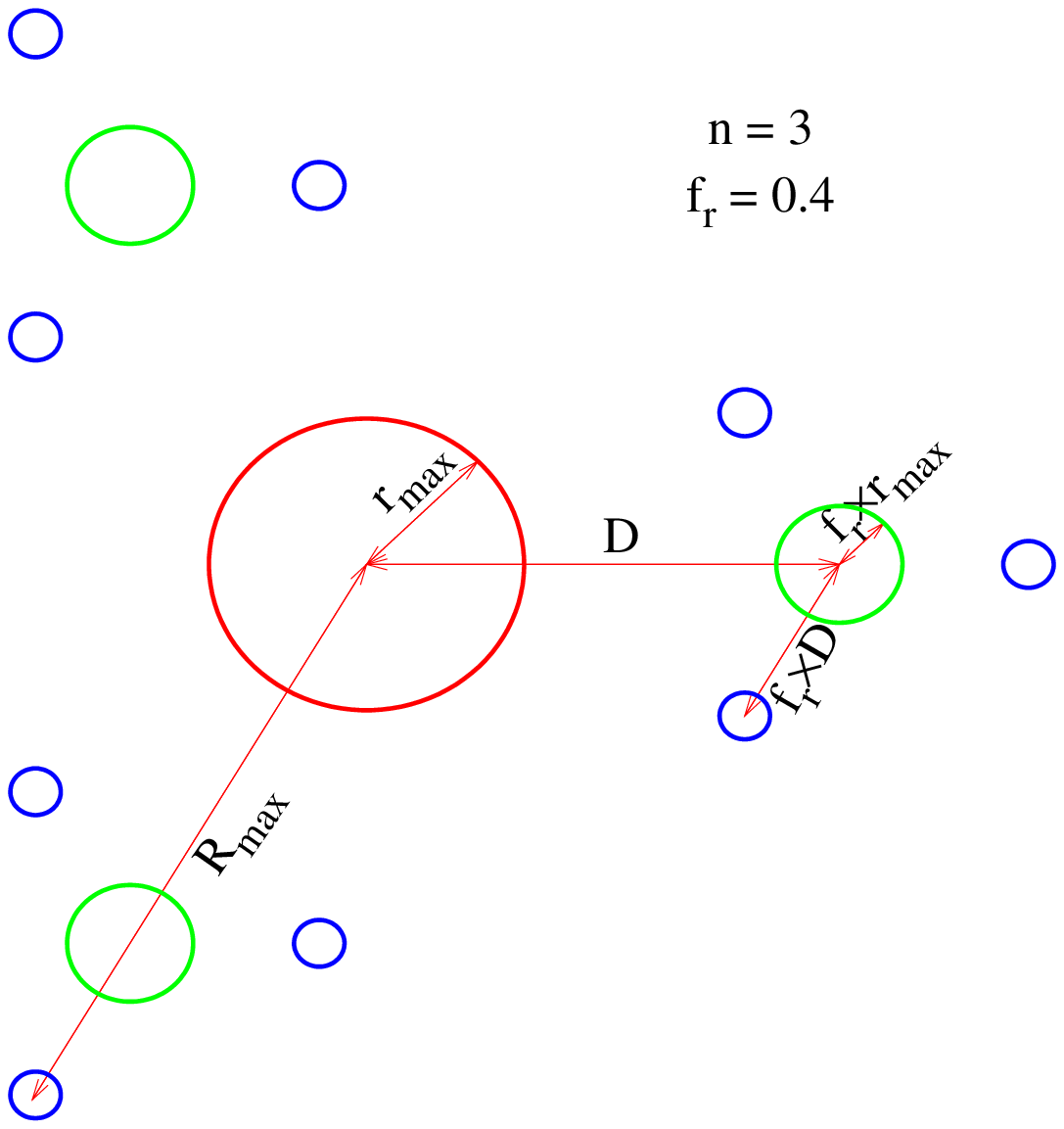}\includegraphics[trim = 25mm 25mm 25mm 25mm, clip, width=6.25cm]{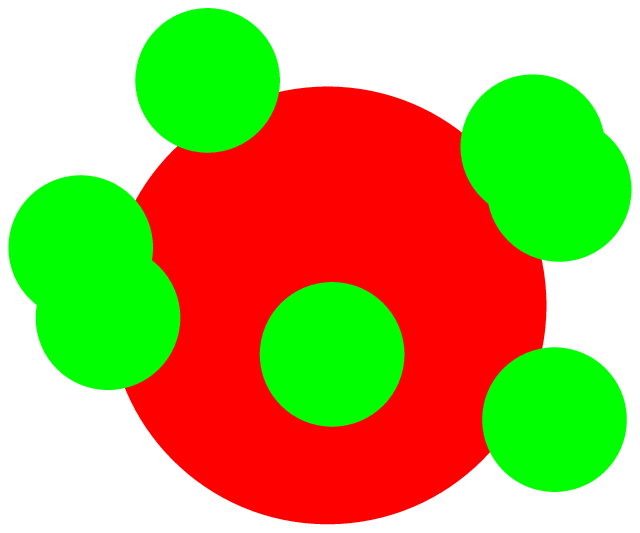}\includegraphics[trim = 25mm 25mm 25mm 25mm, clip, width=6.25cm]{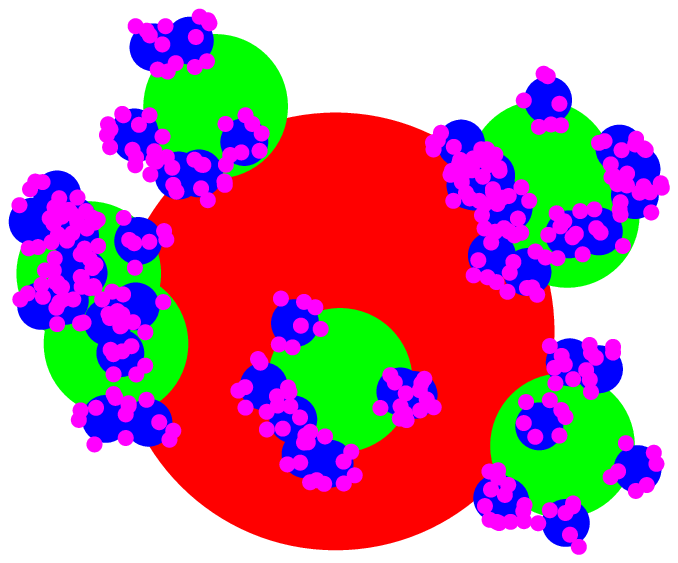}
\caption{\label{fig:dmfrac} {\it Left:} Schematic representation of the halo substructure with three subhalos around any big halo. Note that both the cut-off radius ($r_{max}$) and the halo to subhalo distance ($D$) are scaled down by a factor $f_r$ for each substructure level. {\it Middle and Right} Halo substructure with seven subhalos around any big halo and with $f_r=0.33$. The middle and right panels show the structure with two and four substructure levels respectively.}
\end{center}
\end{figure*}

\subsection{Assumptions, Parameters and Constraints}
\label{sec:param}

Assuming that the dark matter clustering has a scale free nature (i.e., there 
exist halo substructures of a wide range of mass), the density profile can be 
described as a combination of a smooth radial profile $\overline{\rho(r/r_c)}$ 
and a stochastic part $\delta\rho$. Here $r_c$ is a characteristic ``core'' 
radius and $\overline{\rho}$ contains information of the density variation at 
scales larger than or comparable to $r_c$. For the purpose of this work, we 
have used a simpler model of this density distribution which, however, retains 
all relevant key features. In this simplified model, we have assumed that each 
dark matter clump has a number of smaller clumps around it. Each of these 
smaller clumps are in turn just a scaled down version with even smaller clumps 
around them. As a result, the whole structure has an approximate spherical 
symmetry and a statistical self-similarity. It is assumed here that all these 
clumps have non-singular isothermal density profile where the central density 
($\rho_0$), the core radius ($r_c$), the cut-off radius ($r_{max}$) and the 
halo to subhalo distance ($D$) are scaled down accordingly. However, this 
specific density profile is not a crucial assumption in our model and the 
individual substructures may have any non-singular density profile 
$\rho(r/r_c)$, where $r_c$ is some characteristic radius. Essentially, this 
is a fractal structure with three parameters: (i) $n$, the number of small 
clumps around any clump, (ii) $f_r$, spatial scaling factor for core radius, 
cut-off radius and distance and (iii) $f_\rho$, central density scaling 
factor between any clump and its next smallest clumps. A fractal is a 
fragmented and irregular geometrical shape with exact or stochastic 
self-similar structures at all scales \citep{mand83}. Independent of the 
value of $f_r$ and $n$, the (Hausdorff) fractal dimension of such a 
structure is $3$ for $nf_r^3<1$. However, since at each iteration, the linear 
size of clumps scales by a factor $f_r$ and the mass scales by a factor $n$, 
the local mass dimension for any substructure level is 
$D_m = -log(n)/log(f_r)$ over a certain range of scales. A mass dimension of 
$D_m$ for a medium implies that the mass enclosed in a sphere of radius $r$ in 
such a medium will be $M(r) = kr^{D_m}$. So, for the $N^{\rm th}$ substructure 
level, $M_N(r) = k_Nr^{D_m}$ over a range of scales depending on $N$, $n$, 
$f_r$, $D$ and $r_{max}$. Note that this range is different for different 
substructure levels. Hence, for the complete structure, the total mass $M(r)$, 
which is the sum of $M_N(r)$ of all the substructure levels, will not have a 
simple power law radial dependence. But, the dark matter halo mass function 
will still be a power law, $N(m) \propto m^{-\alpha}$, where the power law 
index $\alpha = -log(n)/log(f_r^3 f_{\rho})$ is a more physically motivated 
parameter of this model and can be constrained from theoretical and numerical 
analysis of dark matter structure formation \citep{gao04,zemp09}. The density 
distribution can be written as 
\begin{equation}
\rho(r) = \rho_{\rm bg} + \displaystyle\sum_{i=0}^\infty\sum_{j=1}^{n^i} 
\rho_s(\rho_{0,i}, r_{c,i}, r_{max,i}, \vec{r}_{i,j})
\label{eq:density}
\end{equation}
where $\rho_{\rm bg}$ is background density and $\rho_s(\rho_{0,i}, r_{c,i}, 
r_{max,i}, \vec{r}_{i,j})$ is density profile of individual substructure 
centred at $\vec{r}_{i,j}$ with central density ($\rho_{0,i}$), core radius 
($r_{c,i}$) and cut-off radius ($r_{max,i}$). Considering the self-similarity 
of this model,
\begin{eqnarray}
&&\rho_{0,i} = f_\rho \times \rho_{0,i-1} \nonumber\\
&&r_{c,i} = f_r \times r_{c,i-1} \nonumber\\
&&r_{max,i} = f_r \times r_{max,i} \nonumber\\
&&\vec{r}_{i,j} = \vec{r}_{i-1,k} + (f_r \times D_{i-1} \times \vec{d})
\end{eqnarray}
where $k=n^{i-1}$, $\vec{d}$ is a unit length vector with random orientation 
and the initial set of parameters $\rho_{0,0}, r_{c,0}, r_{max,0} = \rho_{0}, 
r_{c}, r_{max}$ is for the largest subhalo centered at the origin. In 
principle, $\rho_{\rm bg}$ may be a smooth function of $r$. But, since we are 
assuming it to be a small background density threshold, its effect on the final 
rotation curve is not very significant. So, for simplicity, we have assumed 
$\rho_{\rm bg}$ to be constant over the radius of our interest. This structure 
is shown schematically (without any randomness for the shake of clarity) in 
the left panel of Figure~\ref{fig:dmfrac}. After introducing randomness in 
angular position of the subhalos, one realization of such a structure with 
$n=7$ and $f_r=0.33$ is shown with two and four substructure levels in the 
middle and right panel respectively. This model can be considered as a 
simplified representation of the scale free, clumpy density structure of dark 
matter above a small threshold density at the galactic scale.

We note that the parameters for this model are constrained to a good extent by 
various physical considerations. Assuming that the structure is extended down 
to the infinitely small scale, to avoid a divergence of the total mass, the 
quantity $n f_r^3 f_{\rho}$ must be less than unity. Similarly, for a halo 
with cut-off radius $r_{max}$, halo to subhalo distance $D$ must be greater 
than or equal to $(1-f_r)r_{max}/(1-2f_r)$ to avoid any overlap. Also note 
that the density of any subhalo at the cutoff radius $\rho(r_{max})$ scales as 
$f_{\rho}$. So, we adopt $f_{\rho} = 1.0$ for further analysis to ensure a 
constant density at the cutoff radius for all the substructures. In this 
simplified model, we have assumed that all the parameters like $D$, $\rho_0$, 
$r_{max}$ etc.~are identical for all the subhalos for a particular 
substructure level and $n$, $f_r$ and $f_\rho$ remain constant for all the 
levels. In a realistic situation, however, all these parameters may have some 
random variation. As a result of these fluctuations, the halo mass function, 
the radial mass distribution, and in turn, the rotation curves, are expected 
to be somewhat smoother than that are predicted from this analysis.

\subsection{Tidal stability}
\label{sec:tidal}

A much stronger constraint on the parameters comes from the consideration of 
the stability of this structures preventing tidal disruption by invoking 
self-gravity. Considering a rigid object of mass $m$ and radius $r$ at a 
distance $d$ from a bigger object of mass $M$ and radius $R$, from the 
standard Roche limit consideration, $D$ should be $\sim r(2M/m)^{1/3}$ to 
avoid tidal disruption of the smaller body. In the case of any halo and its 
immediate subhalo, $m/M = f_r^3$ and $r/R = f_r$ imply $d \approx 1.25 R$. For 
even smaller subhalo structures with $r/R = f_r^k$, the mass is scaled 
accordingly $m/M = f_r^{3k}$ and $d \approx 1.25 R$ assures stability. Hence, 
a distance
\begin{equation}
D \geq 1.25x\frac{1-f_r}{1-2f_r}r_{max}
\label{eq:roche}
\end{equation}
will make the whole structure stable. Here $x$ is a fudge factor and we use 
the value of $x=1.1$ to accommodate non-rigid density clumps. For a given 
value of $D$, the number of substructure $n$ is also constrained to be
\begin{equation}
n \leq \frac{4\pi D^2}{\pi (f_rD/1-f_r)^2} = \frac{4(1-f_r)^2}{f_r^2}
\label{eq:nsub}
\end{equation}
to avoid any possible overlap of subhalo structures.

\subsection{Virial stability}
\label{sec:virial}

A detailed virial stability analysis requires numerical simulation of the 
dynamics of such a density distribution to get the time-averaged dynamical 
quantities. But, a simple ensemble average virial scaling analysis may be used 
to constrain the central density $\rho_0$ for a set of model parameter. As the 
whole structure is assumed to have an approximate spherical symmetry, average 
kinetic and potential energies, $\langle T \rangle$ and $\langle V \rangle$, 
for thin spherical shells of radius $r$ and thickness $\delta r$ will be
\begin{eqnarray}
&& \langle T \rangle = \frac{1}{2}m\sigma_{\rm DM}^2 = \frac{1}{2} 4\pi r^2\delta r \rho(r) \sigma_{\rm DM}^2 \nonumber\\
&& \langle V \rangle = -\frac{GM(r)m}{r} = -v_c^2(r) 4\pi r^2\delta r \rho(r) 
\end{eqnarray}
where $\sigma_{\rm DM}$ is dark matter velocity dispersion, $M(r)$ is total 
mass within radius $r$ 
and $v_c(r) = (GM/r)^{1/2}$ is the scale dependent 
virial velocity equivalent of the ``circular velocity'' for rotating disk. 
Since the rotation curve has a roughly constant value $v_0$ at large radius, 
the ratio $2\langle T \rangle/|\langle V \rangle|$ will tend to the equilibrium 
value of $1$ at large radius if $\sigma_{\rm DM} \approx v_0$. Using the 
minimum value for $D$ from equation~(\ref{eq:roche}), the maximum extent of 
the structure $R_{max}$ will be $D/(1-f_r)$ and the average density will be
\begin{equation}
\langle\rho\rangle = \frac{3(1-2f_r)^3(s-tan^{-1}s)}{(1.25xs)^3(1-nf_r^3)}\rho_c
\label{eq:vir}
\end{equation}
where $s=r_{max}/r_c$. Note, however, that this is a fractal structure with 
significant porosity. So, the average density of any individual clump is 
higher than $\langle\rho\rangle$ by a factor of 
$(1-nf_r^3)(R_{max}/r_{max})^3$. Now, for global stability of the whole 
structure, $R_{max}$ should be less than or equal to the radius within which 
the virial equilibrium is maintained. Average density within this virial 
radius $r_{vir}$ or $r_{200}$ should be $\approx 200$ times more than the 
critical density $\rho_{crit} = 3H^2/8\pi G$. For a choice of model parameters 
$n$, $f_r$, $s$ and $x$, this will constrain the lower limit of the central 
density $\rho_0$ so that $\langle\rho\rangle \gtrsim 200 \rho_{crit}$. For 
individual substructures, both mass and volume scale as $f_r^3$, keeping the 
average density constant. This implies that stability for one substructure 
level ensures stability for all other levels. 

\section{Results}
\label{sec:result}

\subsection{Rotation curve}
\label{sec:results1}

\begin{figure}
\begin{center}
\includegraphics[width=9cm]{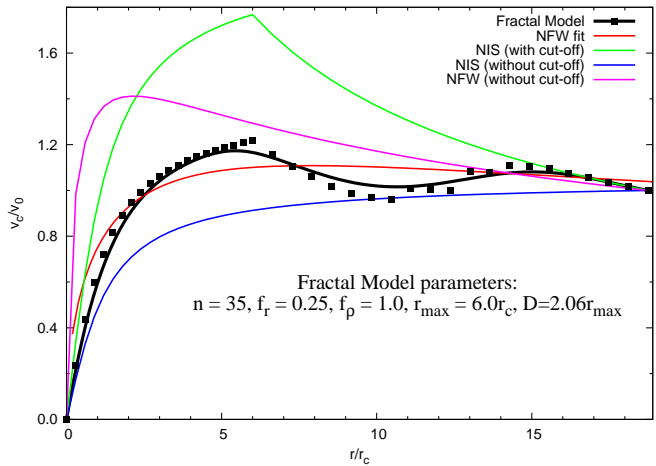}
\caption{\label{fig:rotcur} Predicted rotation curve for the fractal substructure model (black) and the best fit NFW profile to that (red). This is for $n=35$, $f_r=0.25$, $f_{\rho} = 1.0$ and $r_{max} = 6.0 r_c$. Rotation curves for non-singular isothermal sphere with and without a cutoff (green and blue curves respectively) as well as for NFW halo (magenta curve) are also shown for a comparison. Radial distance and rotation velocity are scaled by $r_c$ and $v_0$ (rotation velocity at the furthest radial distance) respectively.}
\end{center}
\end{figure}

Since the halo density distribution is significantly clumpy, the velocity 
field for such a system is also expected to have fluctuations at all scales. 
However, due to approximate spherical symmetry of the clump distribution, the 
average rotation velocity over a spherical shell at radius $r$ will still be 
$\langle v_c(r) \rangle \approx (GM_r/r)^{1/2}$, where $M_r$ is the total mass 
within this radius. As pointed out in subsection \ref{sec:virial} using the 
virial stability argument, the virial velocity or, equivalently, the  
``rotation'' velocity is expected to be approximately same as the local 
velocity dispersion. This derived rotation curve for the fractal model is found 
to be NFW-like at large radial distance. At small radius, by construction, the 
rotation curve is obviously exactly same as that of a non-singular isothermal 
halo. This is shown in Figure~\ref{fig:rotcur}. Radial distance is scaled by 
$r_c$ and rotation velocity is scaled by $v_0$, which is the rotation velocity 
at the maximum radius plotted. The black curve is the predicted rotation curve 
for the fractal substructure model with $n=35$, $f_r=0.25$, $f_{\rho} = 1.0$ 
and $r_{max} = 6.0 r_c$ and the red line is the best fit NFW profile to that. 
The background density threshold $\rho_{bg}$ is assumed to be negligible in 
this model. Rotation curves for non-singular isothermal sphere with and 
without a cutoff (green and blue curves respectively) as well as for NFW halo 
(magenta curve) are also shown in Figure~\ref{fig:rotcur} for a comparison. 
The velocity scaling ensures that the total mass encompassed by the maximum 
radius is same for all models. 

\begin{figure}
\begin{center}
\includegraphics[width=9cm]{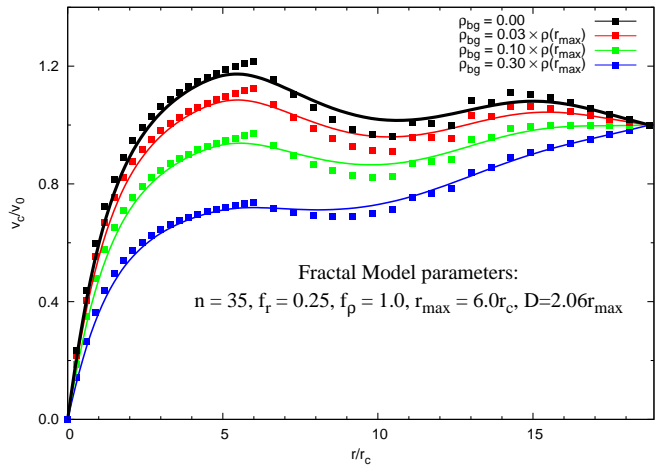}
\caption{\label{fig:rbgcur} Predicted rotation curve for the fractal substructure model for different background density threshold $\rho_{bg}$. All other parameters are same as in Figure~\ref{fig:rotcur}. Different curves are for $\rho_{bg}/\rho(r_{max}) = 0.00$, $0.03$, $0.10$ and $0.30$ (black, red, green and blue curve respectively).}
\end{center}
\end{figure}

The effect of the background density threshold $\rho_{bg}$ is shown in 
Figure~\ref{fig:rbgcur}. Here, the rotation curve is evaluated for different 
$\rho_{bg}$ keeping all other parameters same as in Figure~\ref{fig:rotcur}. 
For scaling, we have used the density $\rho(r_{max})$ at the cutoff radius 
$r_{max} = 6.0 r_c$. Different curves in Figure~\ref{fig:rbgcur} are for 
$\rho_{bg}/\rho(r_{max}) = 0.00$, $0.03$, $0.10$ and $0.30$ (black, red, green 
and blue curve respectively). The general NFW-like nature of the rotation 
curve and the radial fluctuations remain unchanged. But depending on the value 
of $\rho_{bg}$, rotation curve at large radius may be rising, flat or 
declining. Note that the derived rotation curves shown in 
Figure~\ref{fig:rbgcur} are with the simple model of a constant $\rho_{bg}$. 
In reality, $\rho_{bg}$ is expected to be decreasing with increasing $r$, 
giving rise to a rotation curve somewhat intermediate between the extremes 
shown in Figure~\ref{fig:rbgcur}.

\subsection{Velocity fluctuations}
\label{sec:results2}

The predicted rotation curve due to the clustering of dark matter subhalos is 
very similar to the observed rotation curve and the empirical NFW rotation 
curve. However, unlike the NFW model with a smooth radial density distribution, 
the present model predicts a significant fluctuation of rotation velocity in 
both angular and radial directions. Since the underlying density field, which 
gives rise to this velocity fluctuations, is scale free, the velocity 
fluctuation power spectrum is expected to be a power law. Though the fractal 
model has many free parameters, the only relevant parameters for the index of 
this power law are $n$ and $f_r$ while the rest of them will just introduce 
different multiplicative scaling. This prediction can be easily verified from 
high spatial and spectral resolution observation of neutral hydrogen of normal 
galaxies. Note that part of this fluctuations will cancel out for the 
spherically averaged rotation curve and hence it is important to use the full 
velocity field to search for such scale free fluctuations. It is also 
important to note that fluctuations of the velocity field of the hydrogen gas 
will have contributions from the local density perturbations of the disk. But 
the scale dependence of these perturbations may makes it possible to 
disentangle the scale free perturbations due to the halo substructures. There 
is some indication of such a power law scaling of the rotation velocity 
fluctuation power spectra from direct observation and analysis of the 
H~{\sc i} 21~cm velocity field of some nearby galaxy \citep{mdr10}. We leave a 
more detailed treatment of this aspect to a future work.

\section{Discussion}
\label{sec:disc}

Throughout this analysis we have assumed that the clump distribution in this 
model has an approximate spherical symmetry. This assumption is most likely 
to be untrue in reality. Due to small count, bigger substructures are more 
likely to cause departure from spherical symmetry. In this situation, rotation 
velocity field will also have strong azimuthal asymmetry. Observationally, 
about half of the spiral galaxies show some degree of kinematical lopsidedness 
\citep{rs94,hay98,sw99,sofue01,jc09} which may originate from tidal distortion 
partly due to deviation from spherical symmetry of the underlying 
gravitational potential of the lopsided dark matter halo \citep{sc09,saha09}. 
Due to their large number, smaller substructures are likely to have relatively 
more symmetrical distribution. So, the velocity fluctuations spectrum is 
expected to show a power law scaling at large spatial frequency (large $k$, 
i.e., small physical scale), and a departure from power law at small $k$. 

We have also assumed in this analysis that the baryons will not significantly 
affect the galactic dynamics. However, some recent numerical studies 
\citep[e.g.][]{wd08,rd08,rd09} have shown that the presence of baryons have 
effects like flattening the central cusp, reducing the halo triaxiality, and 
introducing bias in clustering with environment density. Effectively, baryon 
dissipation may destroy the similarity between different scales of clustering. 
However, \citet{rd10} have found that phenomenon like ``efficient feedback 
from stellar evolution and the central supermassive black holes'' may 
counterbalance this effect to some extent. On the other hand, \citet{kne10} 
have found no significant effect of baryonic physics on properties like shape 
and radial alignment of substructures. Numerical simulations also suggest 
that, even in the presence of baryons, the subhalo mass function remains a 
power law though the power law index changes from $-0.99$ to $-1.13$ 
\citep{rd10}. It is, hence, important to keep in mind that, in a realistic 
situation, baryon dissipation may alter the velocity fluctuations causing a 
significant departure from the scale free velocity fluctuations.

We note that the key result of this analysis, that is an NFW-like rotation 
curve for the fractal model, is not crucially dependent on the exact density 
profile of individual halos. A variety of density profile without any central 
singularity (variant of non-singular isothermal sphere) will leads to a similar 
NFW-like rotation curve. This strongly suggests that the clustering properties 
of the dark matter particles dominantly govern the radial density profile of 
the halo. Finally, these assembly of substructures will give rise to a flat 
NFW-like rotation curve for normal galaxies. But, for more dark matter 
dominated low mass dwarf galaxies and low surface brightness galaxies, 
depending on the exact form of the non-singular density profile, the dominant 
contribution of the central big halo may make the rotation curve for such 
galaxies intrinsically different which is consistent with observational 
results \citep{blok02,blok05}.

\section{Conclusions}
\label{sec:concl}

\begin{enumerate}
\item The dark matter halo substructures at galactic scale is found to 
significantly affect the rotation curve.
\item A self-consistent model of statistically self-similar hierarchical dark 
matter substructures predicts an NFW-like rotation curve, though each clump 
has a non-singular isothermal density profile. This NFW-like rotation curve 
emerges out of the fractal geometry and is independent of specific density 
profile of individual clumps. 
\item The model predicts a scale free power spectrum of the rotation velocity 
fluctuations which can be observationally verified.
\item The model also provides some plausible explanation of the observed 
intrinsic difference between the dark matter halo density profile of normal 
galaxies and dark matter dominated low mass dwarf galaxies and low surface 
brightness galaxies.
\end{enumerate}

\acknowledgments

We are grateful to Rajaram Nityananda for many useful comments on an earlier 
version of this paper. We thank Susmita Chakravorty, Aritra Basu, Sanjay 
Bhatnagar, Abhirup Datta, Prasun Dutta, Sanhita Joshi and Chandreyee Sengupta 
for helpful discussions. We are also grateful to the anonymous referee for 
useful comments and for prompting us into substantially improving this paper. 
NR is a Jansky Fellow of the National Radio Astronomy Observatory (NRAO). 
This research was supported by the NRAO. The NRAO is a facility of the 
National Science Foundation (NSF) operated under cooperative agreement by 
Associated Universities, Inc. (AUI).

{\it Disclaimer}: This is an author-created, un-copyedited version of an 
article accepted for publication in the Astrophysical Journal. IOP Publishing 
Ltd is not responsible for any errors or omissions in this version of the 
manuscript or any version derived from it. The definitive publisher 
authenticated version will be available online at {\tt http://iopscience.iop.org/}.

\end{document}